\documentclass[aps,prb,reprint,superscriptaddress,longbibliography]{revtex4-1}
\usepackage{amssymb}
\usepackage[utf8]{inputenc}
\usepackage{graphicx}
\usepackage{bm,color,subfigure,amsmath}

\newcommand{\m}{\textbf{m}}

\newcommand{\abs}[1]{\left| #1 \right|}

\begin{document}

\title{Transverse instabilities of stripe domains in magnetic thin films with perpendicular magnetic anisotropy}

\author{Max~E.~Ruth}
\affiliation{Department of Applied Mathematics, University of Colorado, Boulder, Colorado 80309, USA}

\author{Ezio~Iacocca}
\affiliation{Department of Applied Mathematics, University of Colorado, Boulder, Colorado 80309, USA}
\affiliation{Department of Physics, Division for Theoretical Physics, Chalmers University of Technology, 412 96, Gothenburg, Sweden}

\author{Panayotis~G.~Kevrekidis}
\affiliation{Department of Mathematics and Statistics, University of Massachusetts, Amherst, MA 01003, USA}

\author{Mark~A.~Hoefer}
\email{hoefer@colorado.edu}
\affiliation{Department of Applied Mathematics, University of Colorado, Boulder, Colorado 80309, USA}

\begin{abstract}
  Stripe domains are narrow, elongated, reversed regions that exist in
  magnetic materials with perpendicular magnetic anisotropy. Stripe
  domains appear as a pair of domain walls that can exhibit topology
  with a nonzero chirality.  Recent experimental and numerical
  investigations identify an instability of stripe domains in the long
  direction as a means of nucleating isolated magnetic skyrmions.
  Here, the onset and nonlinear evolution of transverse instabilities
  for a dynamic stripe domain known as the bion stripe are
  investigated.  Both non-topological and topological variants of the
  bion stripe are shown to exhibit a long-wavelength transverse
  instability with different characteristic features.  In the former,
  small transverse variations in the stripe's width lead to a neck
  instability that eventually pinches the non-topological stripe into
  a chain of two-dimensional breathers composed of droplet soliton
  pairs.  In the latter case, small variations in the stripe's center
  results in a snake instability whose topological structure leads to
  the nucleation of dynamic magnetic skyrmions and antiskyrmions as
  well as perimeter-modulated droplets.  Quantitative, analytical
  predictions for both the early, linear evolution and the long-time,
  nonlinear evolution are achieved using an averaged Lagrangian
  approach that incorporates both exchange (dispersion) and anisotropy
  (nonlinearity).  The method of analysis is general and can be
  applied to other filamentary structures.
\end{abstract}

\maketitle


\section{Introduction}

Solitons~\cite{Ablowitz1981} are localized structures that are
ubiquitous in nonlinear media such as fiber
optics~\cite{mollenauer2006}, water
waves~\cite{johnson_modern_1997} or atomic condensates~\cite{kev_2015}.
In general, solitons manifest a
balance between nonlinearity and dispersion. Magnetic materials
exhibit both nonlinearity and dispersion associated, respectively,
with anisotropy and exchange. In their simplest manifestation,
solitons in magnetic materials correspond to one-dimensional domain
walls that separate well-defined magnetic states or
domains~\cite{Hubert2009}. Domain wall nucleation and motion can be
manipulated in nanowires, a feature that has led to the proposal of
domain-wall-based technology for oscillators~\cite{Bisig2009},
magnonics~\cite{Hertel2004,Yan2011}, and
computation~\cite{Allwood2002,Parkin2008}. It is also possible for a
pair of domain walls to form a dynamically precessing bound state
known as a magnetic bion~\cite{Kosevich1990,Braun2012}.

Two-dimensional solitons can also exist in magnetic
materials with 
uniaxial anisotropy balancing exchange.  Uniaxial anisotropy is
typically achieved in hard magnetic alloys~\cite{Weller2000} or
ultra-thin film multilayers~\cite{Bruno1989,Broeder1991}, and is
commonly referred to as perpendicular magnetic anisotropy (PMA). In
these materials, it has been possible to observe two-dimensional
solitons such as dissipative
droplets~\cite{Hoefer2010,Mohseni2013,Macia2014,Lendinez2015,Chung2016}
and
skyrmions~\cite{Muhlbauer2009,Yu2010,Jiang2015,Jiang2016,Montoya2017},
finding potential applications as oscillators~\cite{Zhou2015} and
information carriers~\cite{Fert2013,Zhang2015}, respectively.

An important figure-of-merit for magnetic solitons is their stability
to perturbations. Topological concepts~\cite{Braun2012} can be
utilized to predict certain aspects of the stability and dynamics of
magnetic solitons. One-dimensional domain walls can be classified by
their chirality $C$, defined as\cite{Braun2012}
\begin{equation}
  \label{eq:14}
  C = \frac{1}{\pi} \int_{-\infty}^\infty \partial_x \Phi dx,
\end{equation}
where $\Phi$ is the azimuthal angle of the magnetization's in-plane
component.  The chirality $C$ describes the magnetization vector sense
of rotation between two domains.  Note that this definition of
chirality is typically utilized to classify domain walls in planar
ferromagnets whereas here, we are considering uniaxial ferromagnets.
The reason we introduce the definition of chirality in \eqref{eq:14}
that counts the $\pi$ rotations of the in-plane magnetization
component rather than one that counts rotations of the out-of-plane
magnetization component $m_z$ is because the magnetic bion considered
in this work always exhibits a positive and a canceling negative
rotation in $m_z$.  The bion's in-plane chirality \eqref{eq:14}
can be nonzero.

Two-dimensional solitons can be categorized into topological classes
according to their skyrmion number\cite{Braun2012}
\begin{equation}
  \label{eq:15}
  S = \frac{1}{4\pi} \int_{-\infty}^\infty \int_{-\infty}^\infty
  \mathbf{m} \cdot (\partial_x \mathbf{m} \times \partial_y
  \mathbf{m}) dx dy,
\end{equation}
that determines how many times the magnetization texture, defined by
the magnetization vector $\mathbf{m}$, can be mapped onto a
sphere. When the chirality and skyrmion number are zero, the state is
considered non-topological or topologically trivial, and indicates
that such a texture can be smoothly deformed or decays in the presence
of magnetic damping to a spatially homogeneous state.

Magnetic soliton topology yields important information about the
collective behavior of multiple solitons. For example, domain walls
with opposite chirality can annihilate into a trivial, homogeneous
state. Conversely, domain walls with equal chirality are topologically
protected from annihilating. In magnetic materials with PMA, such
domain wall bound states form into dynamically precessing
non-topological and topological bions\cite{Kosevich1990,Braun2012}
that, when extended transversely, are called bion stripes.  In the
absence of an applied field, the in-plane magnetization of a
non-topological bion stripe exhibits a counter-clockwise precessional
frequency that corresponds to a positive sign in conventional spin
dynamics; topological bions exhibit clockwise or negative precessional
frequency. 
In the case of localized, two-dimensional solitons, the additional
degree of freedom also leads to richer behavior. Examples include the
merging or annihilation of droplets~\cite{Maiden2013}, dynamical
skyrmions in the presence of radially symmetric
fields~\cite{Zhou2015}, and perimeter modulations of both
textures~\cite{Lin2014,Xiao2017}.

In realistic experiments and potential applications, topological
protection can be compromised by the geometry of the
system. Two-dimensional materials can be approximately achieved by
utilizing films with a thickness smaller than the exchange length (on
the order of $10$~nm), leading to a near-homogeneous magnetization
along the thickness. However, one-dimensional magnetic materials
require lateral confinement in the form of nanowires that are
typically wider than the exchange length due to fabrication method
limitations. Therefore, effectively one-dimensional magnetic solitons
are prone to transverse dynamics as observed, e.g., in externally
driven domain-wall motion~\cite{Lee2007,Yoshimura2015} and
spin-transfer-torque-driven dissipative droplets~\cite{Iacocca2014} in
nanowires. In a more extreme case, transverse dynamics can be
unstable. Such an instability can be detrimental to domain-wall-based
technologies, but can also be used to nucleate two-dimensional
solitons, as shown both numerically and experimentally for single
skyrmions in a series of recent
works~\cite{Jiang2015,Jiang2016,Heinonen2016,Lin2016,Lui2017}. However,
there is neither an analytical description of such an instability nor
a systematic understanding of the number and topology of resultant
two-dimensional textures.  The transverse instability of
quasi-one-dimensional structures is a subject of interest in its own
right, as it can provide a control handle towards the design of
configurations with a particular number of two-dimensional textures as
has been proposed, e.g., for vortices in atomic Bose-Einstein
condensates~\cite{ma_2010}.  Nonlinear mathematical tools are needed
that can account for both anisotropy and exchange. A particularly
useful method is the average Lagrangian approach~\cite{malomed2002},
where nonlinear dynamics of transverse soliton modulation can be
analyzed. To investigate the stability of transverse dynamics, we
utilize the average Lagrangian approach applied to bion stripes in a
two-dimensional thin film with PMA.

In this paper, we show that bion stripes are transversely unstable in
a manner that depends on their topology. The non-topological bion
exhibits a symmetric or ``neck'' instability that eventually
pinches and nucleates breathers composed of droplet pairs. 
The topological bion exhibits an antisymmetric or ``snake''
instability that nucleates a series of topological defects that evolve
into droplets, skyrmions, and anti-skyrmions. The number of droplets
and topological defects per unit length can be estimated by the most
unstable transverse mode, which enables us to control the dynamical
outcome of our numerical simulations. However, long time dynamics
exhibit soliton interactions that fall outside the applicability of
our analytical approach. The properties of the long-wavelength
transverse instability allow us to determine a nanowire lateral
confinement for which the bion stripe is stable. Our study introduces
a method to analytically describe the nonlinear dynamics of stripe
domains in magnetic materials.

The remainder of the paper is organized as follows. In
Sec.~\ref{sec:analytical-model}, we introduce the analytical model for
magnetization dynamics and the analytical form of a bion stripe
solution. The linear stability analysis of bion stripe transverse
perturbations is studied in Sec.~\ref{sec:bion-filam-stab} using the
average Lagrangian method and numerical linearization. In
Sec.~\ref{sec:nonl-evol-bion}, the nonlinear evolution of a bion
filament is studied and Sec.~\ref{sec:bion-stripe-inst} presents
numerical simulations detailing the filamentary breakup. A discussion
of the implications of our analysis for stabilized bions in physically
confined structures is presented in
Sec.~\ref{sec:disc-bion-strip}. Finally, we provide concluding remarks
in Sec.~\ref{sec:conclusions}.


\section{Analytical model}
\label{sec:analytical-model}

Magnetization dynamics can be analytically described over sufficiently short time scales by the conservative Larmor torque equation~\cite{Kosevich1990}
\begin{equation}
\label{eq:Larmor}
  \partial_t\mathbf{m} = -\mathbf{m}\times\mathbf{h}_\mathrm{eff}=\mathbf{m}\times\frac{\delta\mathcal{E}}{\delta\mathbf{m}},\\
\end{equation}
expressed here in dimensionless form by rescaling time, space, and
fields such that $|\mathbf{m}|=1$. Time is scaled by
$[\abs{ \gamma} \mu_0 M_s (Q-1)]^{-1}$ where $\gamma$ is the
gyromagnetic ratio, $\mu_0$ is the vacuum permeability, and
$Q = 2 K_u/(\mu_0 M_s^2)>1$ where $K_u$ is the uniaxial anisotropy
constant and $M_s$ the saturation magnetization; space is scaled by
$\lambda_{\mathrm{ex}}/\sqrt{Q-1}$ where $\lambda_{\mathrm{ex}}$ is
the exchange length; and fields are scaled by $\sqrt{Q-1}M_s$. The
effective field $\mathbf{h}_{\rm eff}$ includes relevant physics for
the magnetic system studied. Here, we consider a perpendicular
external field $h_0$, exchange field, and perpendicular uniaxial
anisotropy
\begin{equation}
\label{eq:heff}
	\mathbf{h}_\mathrm{eff} = \underbrace{h_0\hat{\mathbf{z}}}_\text{external}+\underbrace{\Delta\mathbf{m}_{ }}_\text{exchange} + \underbrace{m_z\hat{\mathbf{z}}.}_\text{uniaxial anisotropy}
\end{equation}
We assume a sufficiently thin and transversely extended film so that long-range dipolar fields are negligible and the magnetization does not vary through the film thickness, i.e., it is a two-dimensional magnet. The effective field can be described as the functional derivative of the magnetic energy density, $\mathcal{E}$ (defined below), with respect to the magnetization vector.

For the following analysis, it is convenient to represent Eqs.~\eqref{eq:Larmor} and \eqref{eq:heff} in spherical coordinates. For this, we introduce the transformation
\begin{equation}
\label{eq:spherical}
    \textbf{m}(x,y,t) = 
    \left[\begin{array}{c}
        \cos\left[\Phi(x,y,t)+h_0t\right]\sin\left[\Theta(x,y,t)\right]  \\
        \sin\left[\Phi(x,y,t)+h_0t\right]\sin\left[\Theta(x,y,t)\right]  \\
        \cos\left[\Theta(x,y,t)\right]
    \end{array}\right],
\end{equation}
where $\Phi$ is the phase relative to the precession induced by the applied field $h_0$ and $\Theta$ is the polar angle of the magnetization. In this coordinate system, the energy density can be written as~\cite{Hoefer2010}
\begin{equation}
\label{eq:energy}
    \mathcal{E}(\Theta,\Phi) = \frac{1}{2}\left[ \abs{\nabla\Theta}^2 + \sin^2(\Theta)(1 + \abs{\nabla\Phi}^2)\right].
\end{equation}
Note that the perpendicular field has been completely scaled out of
the energy. In other words, any external field along
$\hat{\mathbf{z}}$ represents a frequency shift that is embedded in
the precessional frequency by virtue of the spherical transformation
of Eq.~\eqref{eq:spherical}.

We recast the Larmor equation \eqref{eq:Larmor} as dynamical equations
for $\Phi$ and $\Theta$ via the Euler-Lagrange equations for the
Lagrangian 
\begin{equation}
  \label{eq:lagr}
  L = \int_{\mathbb{R}^2} \left[(1-\cos\Theta)\Phi_t + \mathcal{E}(\Theta,\Phi)\right]\mathrm{d}x\mathrm{d}y.
\end{equation}
The Euler-Lagrange equations are
\begin{subequations}
  \label{eq:LL}
  \begin{eqnarray}
    \label{LL1}
    \partial_t\Theta &=& \frac{\nabla\cdot\left(\sin^2(\Theta)
                         \nabla\Phi  \right)}{\sin\Theta},\\
    \label{LL2}
    \sin(\Theta)\partial_t\Phi &=&
                                   \frac{1}{2}\sin(2\Theta)(\abs{\nabla\Phi}^2
                                   +   1)-\Delta\Theta,
  \end{eqnarray}
\end{subequations}

\begin{figure}
  \includegraphics[width=1\linewidth]{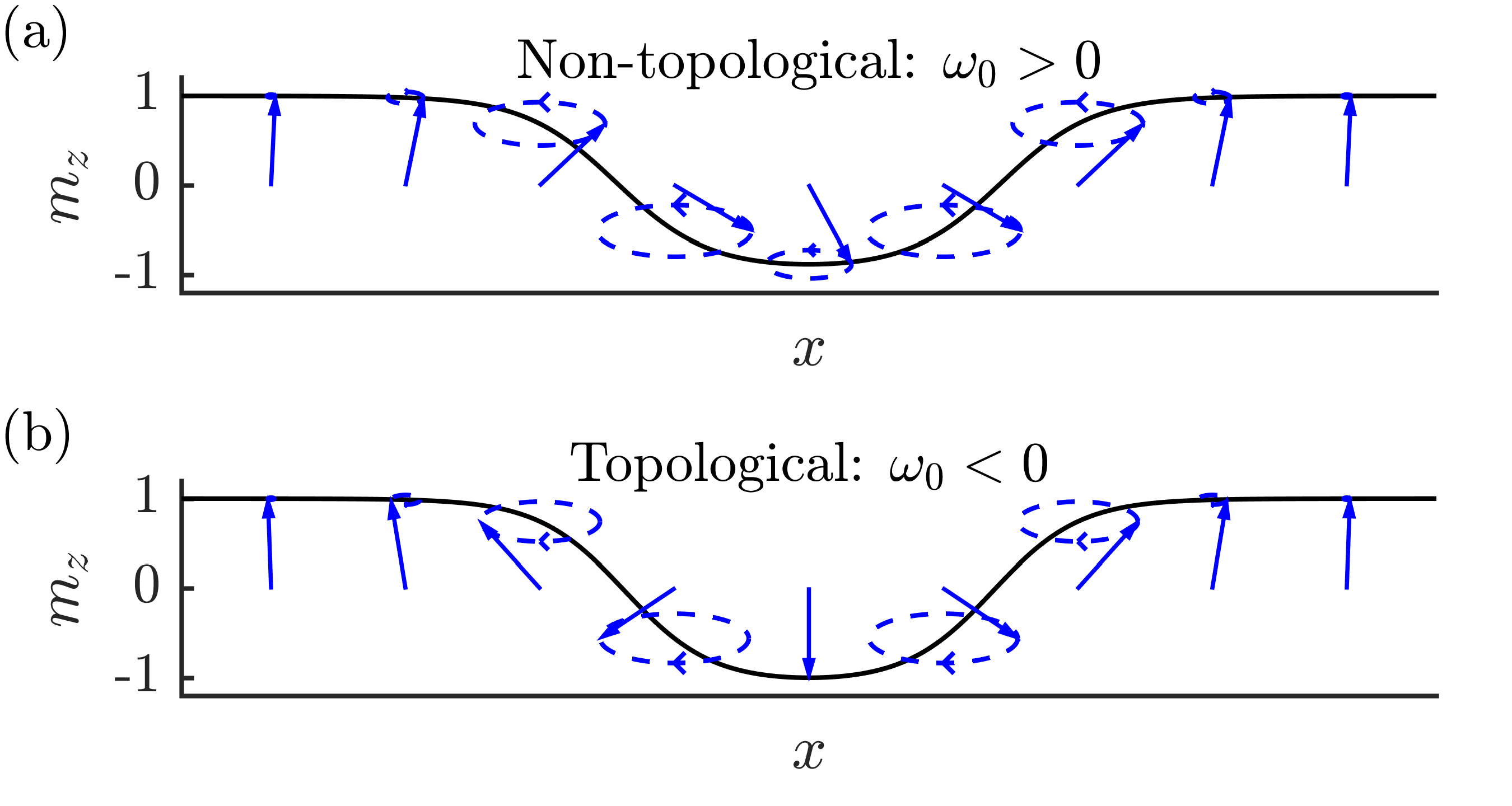}
  \caption{\label{fig:bion} (color online) Profile of the cross
    section of (a) positive frequency, non-topological bion stripe and
    (b) negative frequency, topological bion stripe. The in-plane
    magnetization components and precession direction is schematically
    shown by blue arrows and dashed lines, respectively.}
\end{figure}

An exact solution of the dynamical Eqs.~\eqref{eq:LL} in one
dimension ($\Theta_y = \Phi_y = 0$) is a moving bound state of domain
walls referred to as a bion~\cite{Kosevich1990}. This solution can
trivially be extended to two-dimensions as a bion stripe, which is
expressed as
\begin{subequations}
  \label{eq:bion}
  \begin{eqnarray}
    \label{eq:biontheta}
    &&\cos\Theta(x,y,t) = 1\\ 
    \nonumber
    && \qquad - \frac{4\nu^2}{2-\omega+\sqrt{V^2 +
                          \omega^2}\cosh\left(2\nu[x-\chi(t)]\right)},\\ 
    \label{eq:bionphi}
    &&\Phi(x,y,t) = -\frac{V}{2}[x-\chi(t)]+\phi(t)\\
                &&+ \tan^{-1} \frac{\left[ V^2- 2\left(-\omega +
                    \sqrt{V^2+\omega^2} \right) \right] \tanh \left(
                    \nu[x- \chi(t)]\right)}{2\nu V},\nonumber
  \end{eqnarray}
\end{subequations}
where $\nu = \sqrt{1-\omega - V^2/4}$, $\phi(t) = \omega t + \phi_0$
is the phase including a phase shift $\phi_0$ and $\chi(t) = V t +
\chi_0$ is the center position with an offset $\chi_0$ from the
origin. Therefore, bion stripes are parametrized by four independent
parameters: precessional frequency $\omega$, translational velocity
$V$, phase shift $\phi_0$, and center position shift $\chi_0$. The
precessional frequency $\omega$ is defined relative to the Zeeman
frequency $h_0$. The bion stripe solution of Eq.~\eqref{eq:bion} is
valid for $\omega<1-V^2/4$,~\cite{Kosevich1990} so that $\omega$ can
be positive or negative, which indicates counter-clockwise or
clockwise precession, respectively.

In the limit of small frequency and velocity, $\abs{\omega}\ll1$ and
$\abs{V}\ll1$, the bion stripe behaves similar to a bound state of two
translating domain walls past the Walker
breakdown~\cite{Kosevich1990,Schryer1974}, i.e., translation is
accompanied by magnetization precession about the $\hat{\mathbf{z}}$
component (see Fig.~\ref{fig:bion}).  In the particular case of static
bion stripes, recoverable from Eq.~\eqref{eq:bion} in the limit
$V\to 0^\pm$, the sign of the precessional frequency allows us to
consider two distinct regimes. For positive frequencies, $\omega>0$,
the phase is trivial
$\lim_{V \to 0^\pm} \Phi(x,y,t) = \omega t + \phi_0$ and the chirality
Eq.~\eqref{eq:14} is $C = 0$ so that the stationary bion stripe is a
bound state of domain walls with parallel in-plane phase, shown in
Fig.~\ref{fig:bion}(a). Each domain wall has opposite chirality,
resulting in an overall non-topological state. For negative
frequencies, $\omega<0$, the phase exhibits a $\pi$ jump whose
direction depends on the zero velocity limit
\begin{equation}
  \label{eq:16}
  \lim_{V \to 0^{\pm}} \Phi(x,y,t) = \omega t + \phi_0 \mp \pi
  \mathrm{sgn}\,(x-\chi_0) ,
\end{equation}
so that the chirality evaluates to $C = \mp 1$.  Due to nonzero
chirality, this stationary bion stripe is topological, a bound state
of domain walls with an anti-parallel in-plane phase, shown in
Fig.~\ref{fig:bion}(b).  Note that non-trivial topology implies that
the magnetization at the bion stripe's center is reversed,
$m_z(x=\chi,y,t)=\cos\Theta(x=\chi,y,t)=-1$. The width of the bion
stripe, $\Delta$, is defined as the distance between two crossings of
the magnetization equator, $m_z=0$, and is related to the frequency
and velocity according to
\begin{equation}
  \label{eq:width}
  \Delta = \frac{2 \cosh ^{-1}\left(-\frac{V^2+3 \omega
        -2}{\sqrt{V^2+\omega^2}}\right)}{\sqrt{-V^2-4 \omega +4}}
\end{equation}
Therefore, the domain walls composing a bion stripe approach each other as the frequency or translational velocity increase. In the small frequency and velocity limit, $|\omega|\ll1$ and $|V|\ll1$, the bion stripe width can be approximated as $\Delta = \log\left[4/\sqrt{V^2+\omega^2}\right]$.

Bion stripes offer an analytical probe to study the stability of elongated domains, typical of magnetic materials exhibiting PMA~\cite{Hubert2009}. The topological bion stripe is of particular interest because its structure is reminiscent of chiral N\'{e}el domain walls~\cite{Chen2013,Emori2013} that have been recently utilized to nucleate skyrmions~\cite{Jiang2016,Heinonen2016,Lin2016}. In the following, we will refer to non-topological or topological bion stripes according to their one-dimensional chirality. 


\section{Bion filament stability analysis}
\label{sec:bion-filam-stab}

To study the stability of bion stripes, we determine the evolution of
perturbations along the $\hat{\mathbf{y}}$ direction, i.e., transverse
perturbations.
To attack this nontrivial nonlinear problem from an analytical
perspective, we utilize the average Lagrangian
formalism~\cite{malomed2002} to reduce the dimensionality of the
system. The idea is to assume the modulation of a bion stripe by
allowing its parameters $\omega$, $V$, $\phi$, and $\chi$ to be
functions of $y$ and $t$.  This treats the bion stripe as a soliton
filament or bendable, tube-like curve whose local cross-section is the
bion solution \eqref{eq:bion} that can expand and contract as dictated
by the corresponding Lagrangian (and the resulting Euler-Lagrange
equations).  We remark that another, similar approach to studying the
transverse dynamics of soliton filaments in other areas of nonlinear
physics utilizes an effective Hamiltonian
\cite{kevrekidis_adiabatic_2017}.  By substituting the bion stripe
solution Eq.~\eqref{eq:bion} into the Lagrangian Eq.~\eqref{eq:lagr}
and integrating over $x$, we obtain the averaged Lagrangian. For
simplicity of presentation, we restrict to the low frequency and small
velocity regime where bion stripes approach static stripe domains and
can be topologically classified by the sign of the precessional
frequency. The more general case can be studied in the same manner but
the expressions become more complicated. In the $|\omega|\sim |V|\ll1$
case, asymptotic expansion in frequency, velocity, space, and time
give the leading order averaged Lagrangian (see Appendix
\ref{sec:aver-lagr-bion} for details)
\begin{equation}
  \label{eq:Lave}
  \begin{split}
    &L_\mathrm{avg} = 2\Omega - 2 \partial_Y\left(\phi^2 -
      \chi^2\right) - (\partial_T \phi) \ln(U^2 +
    \Omega^2)\\
    &- \frac{\partial_Y\left(U^2 + \Omega^2\right)}{2(U^2 +
      \Omega^2)} +  4(\partial_T \chi) \tan^{-1} \left(\frac{-\Omega +
        \sqrt{U^2 + \Omega^2}}{U}\right),
  \end{split}
\end{equation}
where $\Omega=\omega/|\omega_0|$, $U=V/|\omega_0|$, $T=|\omega_0|t$,
and $Y=\sqrt{|\omega_0|}y$ denote, respectively, the low frequency,
small velocity, long time, and large space variables and
$|\omega_0|\ll1$ is the characteristic precessional frequency of a
bion stripe.

The averaged equations of motion are the Euler-Lagrange equations of
the averaged Lagrangian Eq.~\eqref{eq:Lave}, which can be expressed in
a symmetric form
\begin{subequations}
  \label{eq:ELave}
  \begin{eqnarray}
    \label{eq:ELave1}    
    \partial_T\chi - \frac{1}{2}\partial_{YY}\alpha - e^u \cos\alpha &=& 0,\\
    \label{eq:ELave2}    
    \partial_T\phi - \frac{1}{2} \partial_{YY}u -  e^u\sin\alpha &=& 0,\\
    \label{eq:ELave3}
    \partial_T\alpha+2\partial_{YY}\chi &=& 0,\\
    \label{eq:ELave4}
    \partial_Tu + 2\partial_{YY}\phi &=& 0,
  \end{eqnarray}
\end{subequations}
with the change of variables
\begin{eqnarray}
    U + i \Omega = e^{u+i\alpha}.
\end{eqnarray}
Equations~(\ref{eq:ELave1}-d) approximate the transverse dynamics of
the bion stripe as a soliton filament and are the primary result of
this work.
It is important to note that, in deriving the
system~(\ref{eq:ELave1}-d), we have retained the nonlinear character
of the problem. Thus, while the underlying linearized dynamics of the
stripe are implicit within this formulation, Eqs.~(\ref{eq:ELave1}-d)
are in principle able to follow the system beyond the stage of
linearized evolution.  We note that Eqs.~\eqref{eq:ELave} can be
rewritten in a mathematically elegant form in terms of the scalar,
complex-valued quantity $Z = u + i \alpha = \ln ( U + i \Omega)$:
\begin{equation}
  \label{eq:11}
  Z_{TT} + Z_{YYYY} + 2 i \left ( e^{\overline{Z}} \right )_{YY} = 0 ,
\end{equation}
where $\overline{Z} = u - i \alpha$ is the complex conjugate.  In
principle, Eqs.~(\ref{eq:ELave1}-d) can be derived using an
alternative, multi-scale asymptotic and differential geometry approach
as was done for dark soliton stripe dynamics in the two-dimensional
nonlinear Schr\"{o}dinger equation~\cite{mironov_dynamics_2011}.
While there are some advantages to using the intrinsic variables of
the latter formulation such as the arc-length and normal-to-curve
spatial variables as independent variables and the curvature of the
filament as one of the dependent variables, we will not pursue this
approach here.

Equations~(\ref{eq:ELave1}-d) with
$U_0 + i \Omega_0 = e^{u_0 + i \alpha_0}$ admit the exact solution
\begin{equation}
  \label{eq:star}
  \begin{split}
    \Omega(Y,T) = \Omega_0 = \mathrm{sgn} \left(\omega_0\right),
    &\quad \chi(Y,T) = U_0T,  \\
    U(Y,T) = U_0, & \quad\phi(Y,T)=\Omega T,
  \end{split}
\end{equation}
representing an unperturbed bion stripe. We begin our study of
Eqs.~(\ref{eq:ELave1}-d) with a linear stability analysis. For this,
we linearize Eqs.~(\ref{eq:ELave1}-d) about the bion stripe solution
Eq.~\eqref{eq:star} in the form
\begin{equation}
  \label{eq:lpde}
  \begin{bmatrix}\phi\\ \alpha\\ \chi\\ u\end{bmatrix}\left(Y,T\right)=\begin{bmatrix}\Omega_0T\\ \alpha_0\\ U_0T\\ u_0\end{bmatrix}+e^{iKY+\Lambda T}\begin{bmatrix}\phi_1\\ \alpha_1\\ \chi_1\\ u_1\end{bmatrix} + \mathrm{c.c.}
\end{equation}
where the subscript $1$ indicates a small amplitude and
c.c.~represents the complex conjugate of the previous term. The form
of the sought solution in Eq.~\eqref{eq:lpde} corresponds to a
sinusoidal variation of the bion stripe in the transverse, $Y$,
direction with wavenumber $K$ and exponential temporal growth with
growth rate $\Lambda$.  The linearization of Eq.~\eqref{eq:ELave} with
Eq.~\eqref{eq:lpde} yields four eigenvalues, of which only one,
$\Lambda(K)$, has positive real part
\begin{equation}
  \label{eq:1}
  \Lambda(K) = K \left ( -K^2 + 2 \sqrt{U_0^2 + \Omega_0^2} \right )^{1/2} .
\end{equation}
This positive growth rate for $0 < K < (2\sqrt{U_0^2 +
  \Omega_0^2})^{1/2}$ implies that the bion stripe suffers from a
long-wavelength transverse instability.  The nature of the instability
can be determined by the eigenvector associated with this eigenvalue,
which can be written
\begin{equation}
  \label{eq:2}
  \begin{bmatrix}\phi_1\\ \alpha_1\\ \chi_1\\ u_1\end{bmatrix} =
  \begin{bmatrix}
    U_0 \Lambda(K) \\ 2K^2(\sqrt{U_0^2 + \Omega_0^2} - \Omega_0) \\
    (\sqrt{U_0^2 + \Omega_0^2} - \Omega_0) \Lambda(K) \\ 2U_0 K^2
  \end{bmatrix} .
\end{equation}
Equations \eqref{eq:1} and \eqref{eq:2} yield significant information
about both the early development and late stage of the transverse
instability.  The eigenvector \eqref{eq:2} leads to important
differences in the nature of the instability of non-topological and
topological bion stripes.

\begin{figure}
\includegraphics[width=1\linewidth]{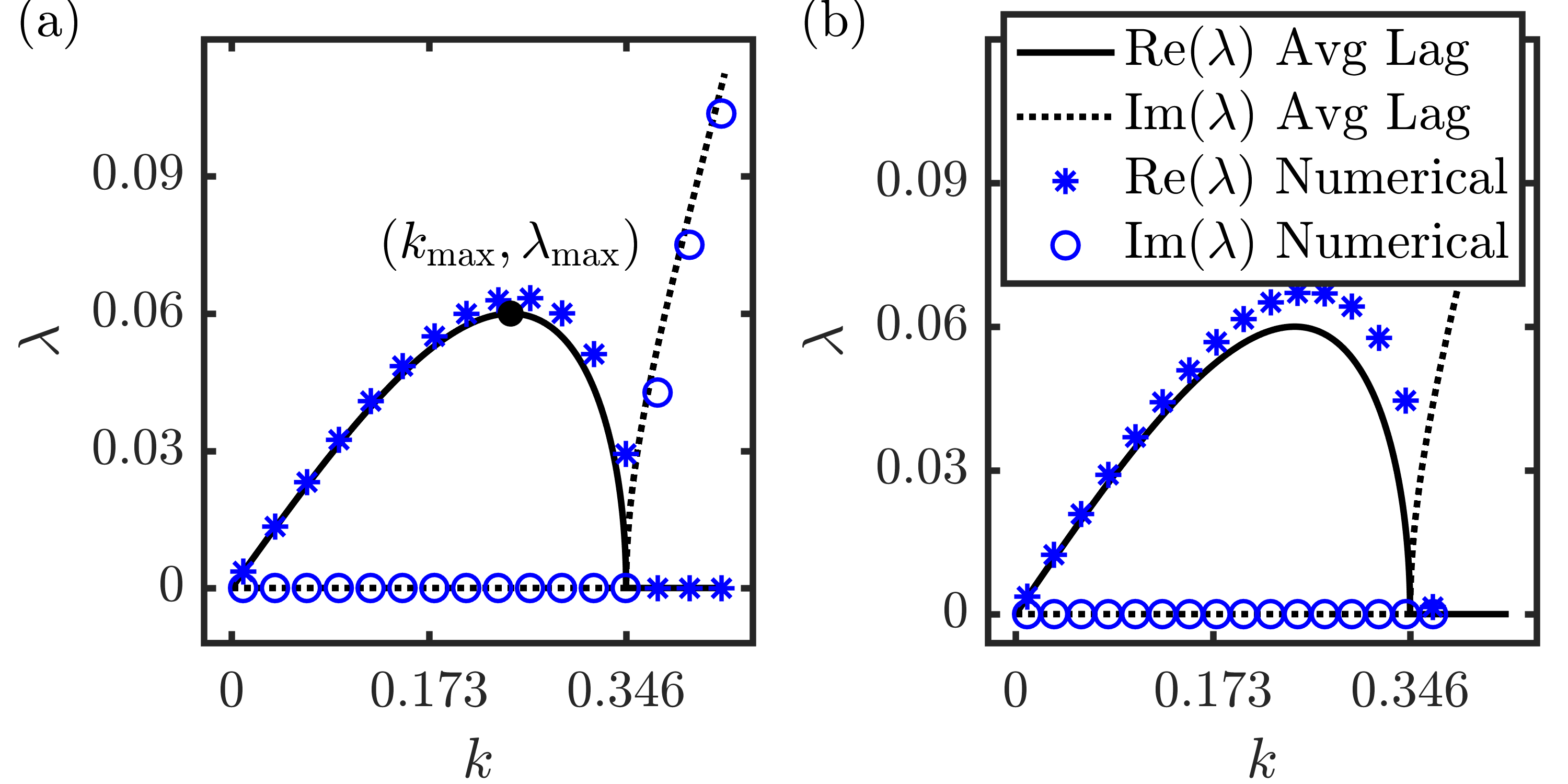}
\caption{\label{fig:dispRel} (color online) Growth rates for the (a)
  non-topological $\omega_0 = 0.06$ and (b) topological $\omega_0 =
  -0.06$ bion stripes. The maximally unstable wavelength and maximal growth
  rate, $k_\mathrm{max}$ and $\lambda_\mathrm{max}$, are indicated by
  a filled black circle in (a). Numerical calculations are shown by blue
  asterisks and circles for the real and imaginary growth rates,
  respectively.}
\end{figure}

\begin{figure}
\includegraphics[width=1\linewidth]{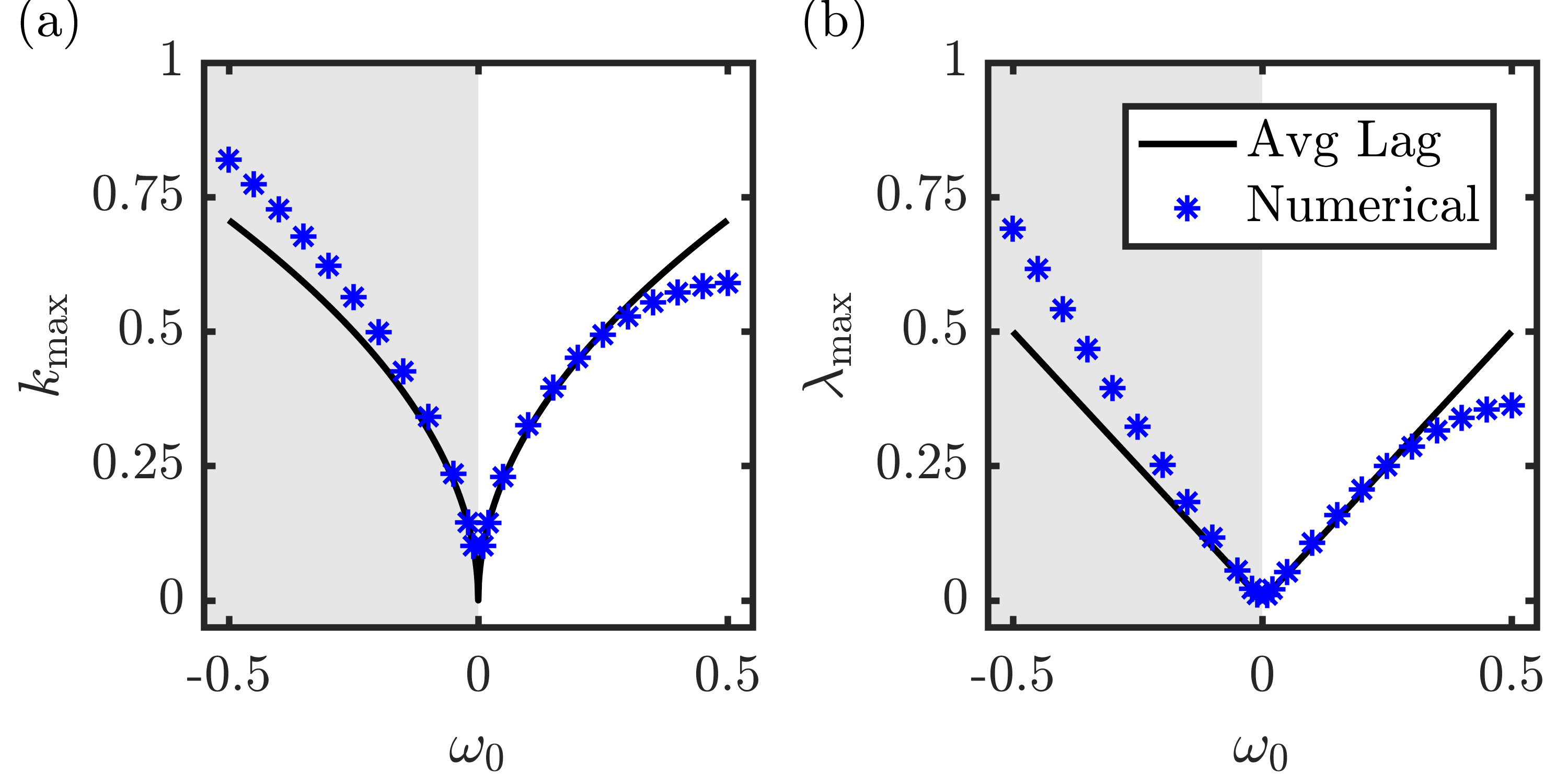}
\caption{\label{fig3} (color online) (a) Maximally unstable wavenumber
  $k_{\rm max}$ and (b) maximal growth rate $\lambda_{\rm max}$ as a
  function of the bion stripe frequency $\omega_0$. The analytical
  calculations are shown by solid black lines and the numerical
  calculations are shown by blue asterisks. The gray and white
  background indicate topological and non-topological bion stripes,
  respectively.}
\end{figure}

\begin{figure}
\includegraphics[width=\linewidth]{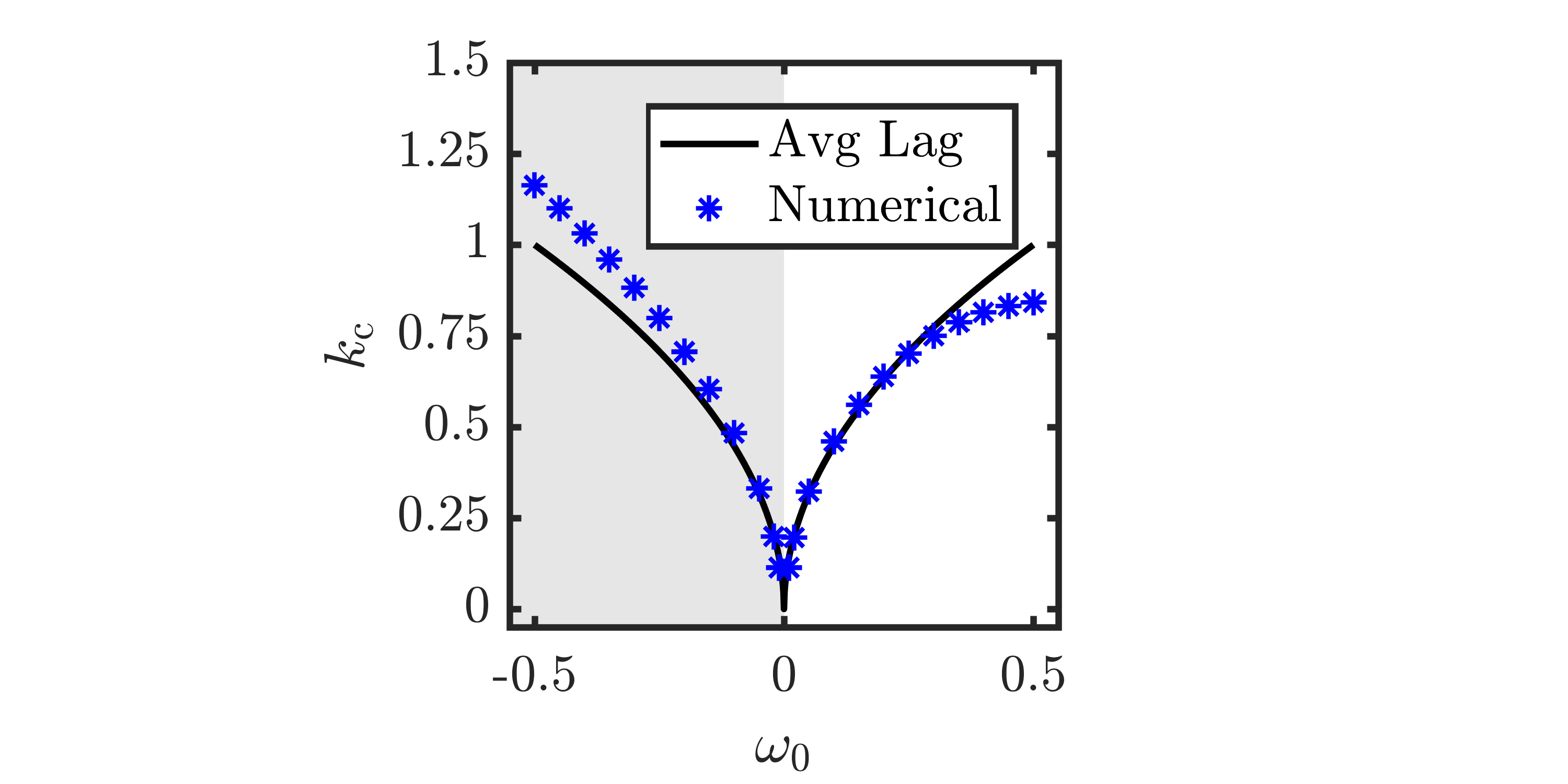}
\caption{\label{fig:kcut} (color online) Unstable band cutoff $k_{\rm
    c}$ as a function of the stationary bion frequency $\omega_0$,
  defined according to $\lambda(k_{\rm c}) = 0$ where $k_{\rm c} >
  0$. The gray and white background indicate topological and
  non-topological bion stripes, respectively.}
\end{figure}

\begin{figure}
\includegraphics[width=1\linewidth]{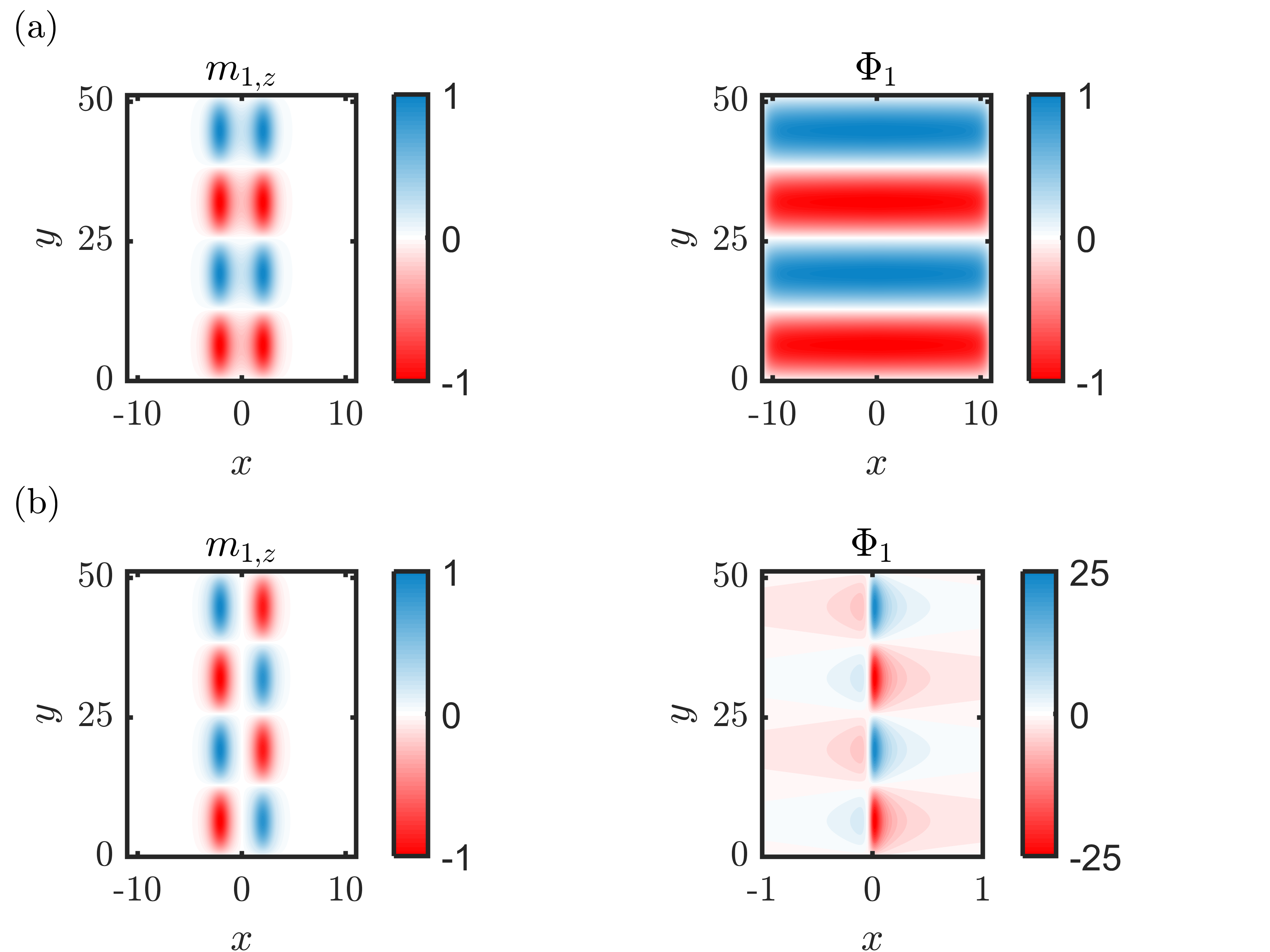}
\caption{\label{fig4} (color online) The computed deviations from the uniform bion stripe, $m_{1,z}$
  and $\Phi_1$ of the maximally unstable modes. 
	(a) Non-topological case
  $\omega_0 = 0.06$ exhibiting the neck instability.  (b) Approximate
  topological case with $\omega_0 = -0.06$, $V = 0.006$ exhibiting the
  snake instability.}
\end{figure}
Our focus in this work is on stationary bion stripes for which
$U_0 = 0$.  In this case, the growth rate \eqref{eq:1} becomes
\begin{equation}
  \label{eq:3}
  \Lambda(K) = K \sqrt{-K^2 + 2},
\end{equation}
because $\Omega_0 = \pm 1$.  All perturbations with wavenumber $K$ in
the unstable band $(0, K_{\rm c})$, $K_{\rm c} = \sqrt{2}$, lead to a
transverse instability. The growth rate \eqref{eq:3} is maximized for
the wavenumber $K_{\rm max} = 1$ and attains the maximal growth rate
$\Lambda_{\rm max} = 1$.  Returning to unscaled variables, the
maximally unstable wavenumber, maximal growth rate, and unstable
wavenumber band for a stationary bion stripe with frequency $\omega_0$
are
\begin{equation}
\label{eq:kmax}
k_{\rm max} = \sqrt{\abs{\omega_0}},\quad\lambda_{\rm max}=
|\omega_0|, \quad k_{\rm c} = \sqrt{2|\omega_0|}.
\end{equation}
The dominant growth rate, wavelength of instability, and unstable band
are the same for topological ($\omega_0 < 0$) and non-topological
($\omega_0 > 0$) bion stripes.

We have also performed a linearization of the Larmor torque equation
\eqref{eq:Larmor}, \eqref{eq:heff} about the bion stripe
solution~\eqref{eq:bion}.  This leads to a linear eigenvalue problem for small
perturbations of the magnetization vector.  Direct numerical
computation yields a definitive prediction for the unstable mode and its
growth rate dependence on the transverse wavenumber $k$.  The details
are described in Appendix~\ref{sec:line-larm-torq}. We use these
numerical computations to verify the usefulness of Eqs.~\eqref{eq:3} and
\eqref{eq:kmax} in capturing the relevant spectrum.
Figure \ref{fig:dispRel} shows the unstable
eigenvalue as a function of the transverse wavenumber from both
Eq.~\eqref{eq:3} rescaled (solid curve) and numerical computation
(asterisks).  For fixed initial bion frequency $\omega_0$, the
non-topological, Fig.~\ref{fig:dispRel}a, and topological,
Fig.~\ref{fig:dispRel}b, growth rates follow the trend predicted by
Eq.~\eqref{eq:3}.  The maximally unstable wavenumber $k_{\rm max}$ and
maximal growth rate $\lambda_{\rm max}$ are shown in Fig.~\ref{fig3}a
and Fig.~\ref{fig3}b, respectively, for a range of initial bion
frequencies.  Finally, the unstable wavenumber band $k_{\rm c}$ is
shown in Fig.~\ref{fig:kcut} from both the average Lagrangian theory,
Eq.~\eqref{eq:kmax}, and numerics.  Figures \ref{fig3} and
\ref{fig:kcut} demonstrate quantitative agreement between 
the instability parameters 
determined by the average Lagrangian theory and Eq.~\eqref{eq:kmax} when $|\omega_0|$ is sufficiently small.  However, for sufficiently large $|\omega_0|$,
deviations are observed and the discrepancies are not symmetric in
$\omega_0$.  In other words, the topological or non-topological
character of the bion influences the instability.

The unstable eigenvector \eqref{eq:2} determines the nature of the
transverse instability and its topological dependence.  We consider
each case in turn.  First, when $\Omega_0 = 1$ (non-topological), we
can divide the eigenvector \eqref{eq:2} by $U_0$ and take the
limit $U_0 \to 0$ to obtain
\begin{equation}
  \label{eq:4}
  \Omega_0 = 1: \qquad \begin{bmatrix}\phi_1\\ \alpha_1\\ \chi_1\\
    u_1\end{bmatrix} = 
  \begin{bmatrix}
     \Lambda(K) \\ 0 \\ 0 \\ 2 K^2
  \end{bmatrix} ,
\end{equation}
for wavenumbers in the unstable band $K \in (0,K_{\rm c})$.  The
nonzero components of the eigenvector determine which bion parameters
exhibit exponential growth.  Evaluating the eigenvector at the maximal
growth rate $\Lambda_{\rm max} = 1$ and associated wavenumber $K_{\rm
  max} = 1$ while assuming an initial perturbation of small amplitude
$a$ in this unstable direction, we find that the bion phase and
frequency exhibit exponential temporal growth
\begin{equation}
  \label{eq:5}
  \phi(Y,T) \sim T + \frac{a}{\sqrt{3}} e^T \cos Y, ~ \Omega(Y,T) \sim 1 + \frac{2 a}{\sqrt{3}}
  e^T \cos Y,
\end{equation}
whereas the bion center $\chi(Y,T) = 0$ and velocity $U(Y,T) = 0$ do
not.  This implies that the topological bion exhibits a transverse
instability whose initial development is dominated by fluctuations in
the bion's phase and frequency.  Because the bion width $\Delta$
[recall Eq.~\eqref{eq:width}] depends on the local bion frequency, we
expect to see the development of fluctuations in $\Delta(Y,T)$ during
the initial stage of the transverse instability with negligible
variation in the soliton filament's center $\chi(Y,T)$.  This is known
as a neck transverse instability~\cite{skryabin_98}.

We also investigate the nature of the transverse instability in the
topological case $\Omega_0 = -1$ by dividing the eigenvector
\eqref{eq:2} by 2 and setting $U_0 = 0$ to obtain
\begin{equation}
  \label{eq:6}
    \Omega_0 = -1: \qquad \begin{bmatrix}\phi_1\\ \alpha_1\\ \chi_1\\
    u_1\end{bmatrix} = 
  \begin{bmatrix}
     0 \\ 2K^2 \\ \Lambda(K) \\ 0
  \end{bmatrix} .
\end{equation}
If we perturb in the most unstable direction \eqref{eq:6}, this time
the exponential growth occurs in the bion center and velocity
\begin{equation}
  \label{eq:7}
  \chi(Y,T) \sim \frac{a}{\sqrt{3}} e^T \cos Y, ~ U(Y,T) \sim
  \frac{2a}{\sqrt{3}} e^T \cos Y,
\end{equation}
while the phase and frequency are stationary $\phi(Y,T) \sim -T$,
$\Omega(Y,T) \sim -1$ for a perturbation amplitude $0 < a \ll 1$.  The
growth of variation in the topological bion's center is called a snake
instability; see, e.g.~Ref.~\onlinecite{mhba_16}, for a recent
discussion.

From the numerical calculations, we have also obtained the spatial
eigenfunctions for the unstable modes. The eigenfunctions are
indicated with the subscript 1 and represent deviations from the
uniform bion stripe.  Figure~\ref{fig4} shows the maximally unstable
mode in the non-topological, Fig.~\ref{fig4}a, and topological,
Fig.~\ref{fig4}b, cases. The structure of the unstable mode coincides
with the predictions from the average Lagrangian theory.  In
particular, the $m_{1,z}$ and $\Phi_1$ modes are in-phase.  The
non-topological case exhibits a symmetric mode that, when added to the
bion, leads to a periodic reduction and increase in the bion's width,
manifesting a precursor of the neck instability.  In the topological
case, the mode is antisymmetric and, when added to the bion, leads to
a periodic shift from left to right of the bion's center, suggesting
the onset of the snake instability.  We were unable to perform a
direct linearization of the topological bion stripe because of its
phase jump at $x = \chi(t)$.  Instead, we linearized non-topological,
propagating bions with $\omega < 0$ and $0 < V \ll 1$.  As $V$ is
decreased, we observe numerical convergence of the unstable eigenvalue
and the associated antisymmetric eigenfunction $m_{1,z}$.  The limit
$V \to 0^+$ is a ``topological limit'' in that the result is a
solution with a jump in the phase $\Phi$; see Eq.~\eqref{eq:16}.  The
bion solution \eqref{eq:bion} with small but nonzero $V$ smooths the
phase jump. This manifests in the numerical linearization by a large
relative amplitude between the eigenfunction component $\Phi_1$ in
comparison with the amplitude of $m_{1,z}$; see Fig.~\ref{fig4}(b).

This linear stability analysis predicts that the early stage of the
transverse instability is dominated by either an increase in $\phi$
and $\Omega$ (non-topological case) or $\chi$ and $U$ (topological
case).  Because the governing equations (\ref{eq:ELave}) are
nonlinear, we expect that at later stages of evolution, the two
growing soliton filament parameters will couple to the other two and
significantly influence the evolution.  We now investigate this more
thoroughly.

\section{Nonlinear evolution of a bion filament}
\label{sec:nonl-evol-bion}

In the previous section, we analyzed the linear stage of the bion
stripe instability.  However, both the neck and snake instabilities
grow in time and eventually deviate from the linearized motion. In
this section we numerically integrate the nonlinear average Lagrangian
Eqs.~(\ref{eq:ELave1}-d) and compare them with direct numerical
simulations of the Larmor torque equation (\ref{eq:Larmor}).  The
Larmor equation simulations are described in
Sec.~\ref{sec:bion-stripe-inst}.

\begin{figure}
\includegraphics[width=1\linewidth]{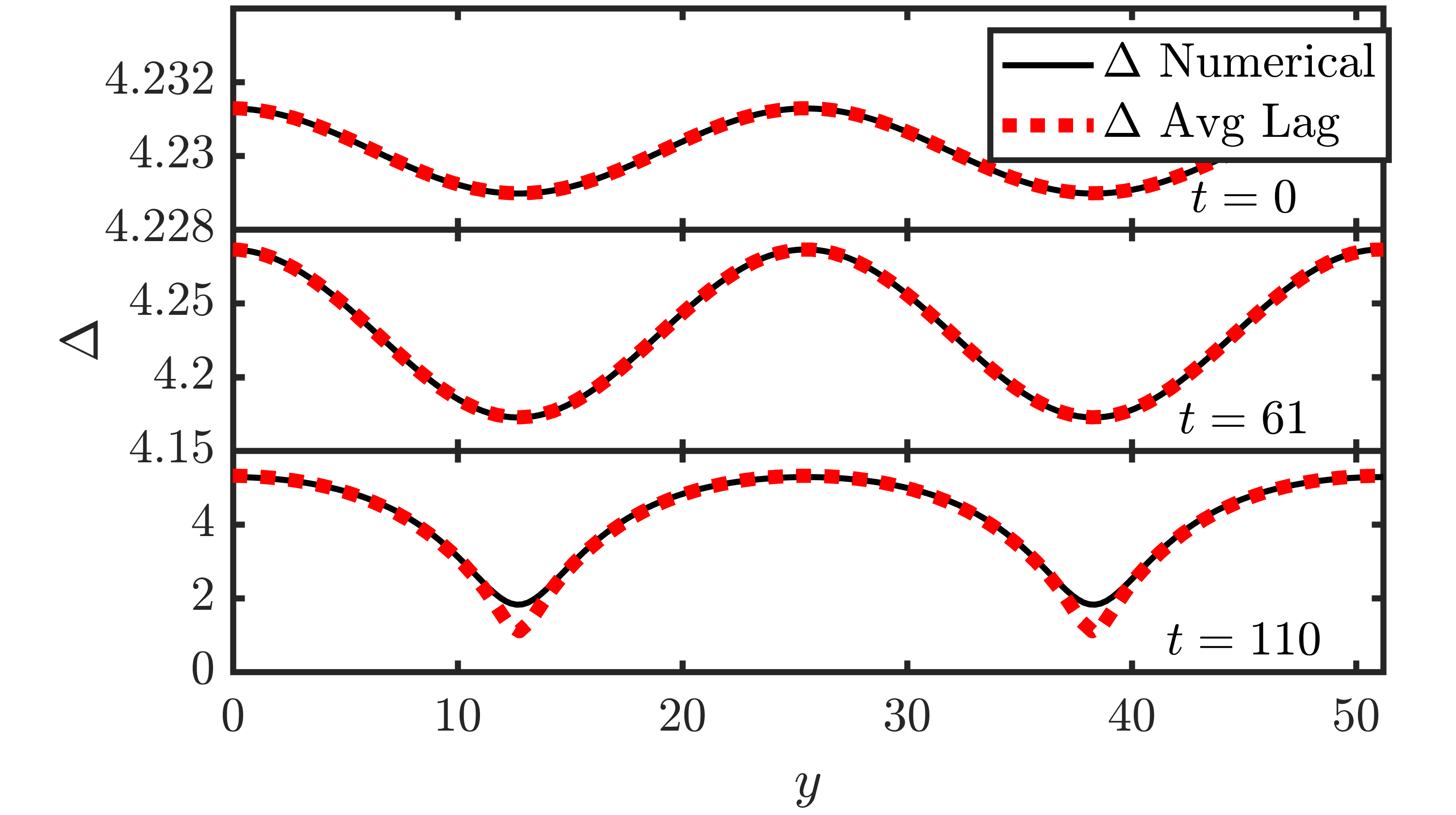}
\caption{\label{fig:NLevolution_nontopological} Evolution of the width
  $\Delta$ of an initially perturbed, non-topological bion stripe with
  $\omega_0 = 0.06$. Direct numerical simulations of the Larmor torque
  equation (solid) and the average Lagrangian equations (dashed) show
  excellent agreement.  The timescale for the average Lagrangian
  results have been scaled by the ratio
  $\lambda_{\rm max}/\lambda_0 \approx 1.06$ where
  $\lambda_0$ is the maximum growth rate from numerical
  linearization.}
\end{figure}
We numerically solve the soliton filament equations (\ref{eq:ELave})
using a pseudospectral discretization in $Y$ and standard fourth order
Runge-Kutta timestepping in $T$. In the non-topological case, we
initialize the soliton parameters with the exact solution
(\ref{eq:star}) plus a small perturbation in the maximally unstable
direction (\ref{eq:4}) with amplitude $a = 10^{-3}$ and sinusoidal $Y$
variation with wavenumber $k_{\rm max}$.  Figure
\ref{fig:NLevolution_nontopological} shows the evolution of the
soliton width parameter $\Delta$ [Eq.~(\ref{eq:width})] (dashed).  As
shown in Fig.~\ref{fig3}b, there is a small discrepancy between the
predicted maximum growth rate $\lambda_{\rm max} = |\omega_0|$ and the
computed maximum real eigenvalue from numerical linearization
$\lambda_0(\omega_0)$. In order to compare the evolution of the
soliton width with direct numerical simulations of the Larmor torque
equation, we have rescaled time in
Fig.~\ref{fig:NLevolution_nontopological} by the ratio of these growth
rates $\lambda_{\rm max}/\lambda_0$.  The Larmor simulations are
initialized with a bion stripe with $\omega_0=0.06$ plus the same
perturbation as the non-topological averaged Lagrangian numerics, with
the frequency perturbation scaled by $a\omega_0$.  The width is
extracted from Larmor simulations by interpolating the numerical
solution to find $x_-(y,t) < x_+(y,t)$ such that $m_z(x_\pm,y,t) = 0$.  The
width reported in Fig.~\ref{fig:NLevolution_nontopological} (solid
curves) is $x_+-x_-$.  The average Lagrangian equations are in
excellent agreement with the full Larmor torque equation, even well
beyond the linear regime.

In Fig.~\ref{fig:NLevolution_nontopological}, we observe significant
amplitude growth and deviation from a sinusoidal waveform to one in
which the soliton width approaches zero, the neck instability.  Zero
width corresponds to pinching of the soliton filament and the
breakdown of the single soliton filament approximation.  The soliton
filament center $\chi$ remains at zero throughout the
simulation. Longer evolution leads to a significant increase in the
frequency $\Omega$, beyond the regime of validity,
$\Omega = {\cal O}(1)$, and therefore signals the breakdown of the
average Lagrangian approach.  We will investigate the pinching of the
soliton filament and subsequent evolution in
Sec.~\ref{sec:bion-stripe-inst}.

\begin{figure}
\includegraphics[width=1\linewidth]{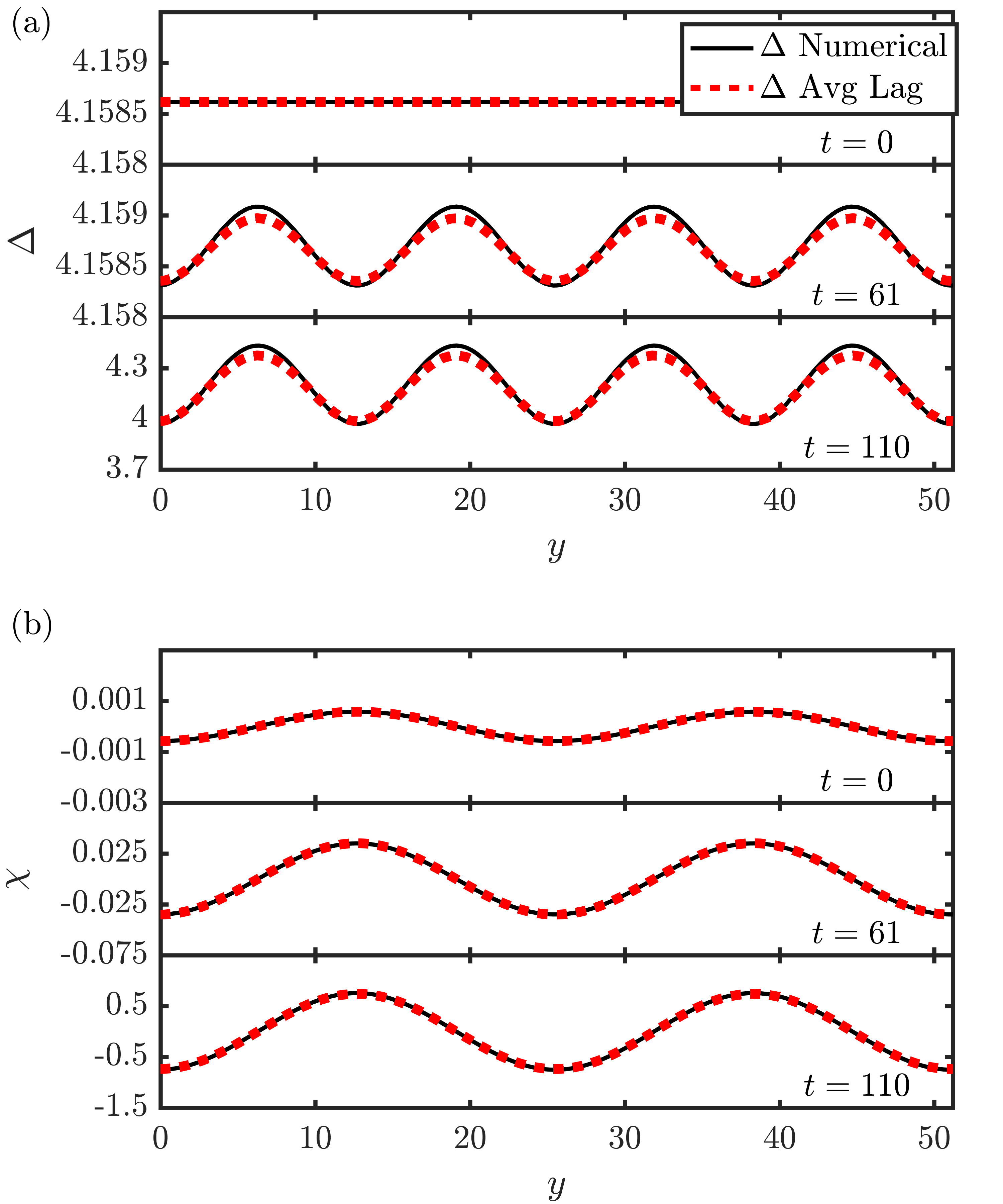}
\caption{\label{fig:NLevolution_topological} Numerical evolution of a
  perturbed, topological bion stripe with $\omega_0 = -0.06$ according
  to the average Lagrangian equations (dashed) and the Larmor torque
  equation (solid). (a) The soliton filament width $\Delta$.  (b) The
  soliton filament center $\chi$.  For both, the timescale for the
  average Lagrangian results have been scaled by the ratio
  $\lambda_{\rm max}/\lambda_0 \approx 1.11$ where $\lambda_0$ is the
  maximum growth rate from numerical linearization.}
\end{figure}

We now investigate the nonlinear stage of evolution of the topological
bion filament. Figures \ref{fig:NLevolution_topological}a and
\ref{fig:NLevolution_topological}b display the evolution of the
soliton filament width $\Delta$ and center $\chi$, respectively, from
numerics of both the average Lagrangian equations (dashed) and the
Larmor torque equation (solid).  Again we rescale time in these
figures by $\lambda_{\rm max}/\lambda_0$ according to the small
difference in the maximal growth rates.  Here, the average Lagrangian
equations (\ref{eq:ELave}) are initialized with a stationary
topological bion perturbed in the maximally unstable direction
(\ref{eq:6}) with amplitude $a = 10^{-3}$.  The Larmor torque equation
is initialized with a bion stripe with frequency $\omega_0 = -0.06$
and the same sinusoidal perturbation, now with the $V$ component
scaled by $\abs{\omega_0}$.  The initially small soliton filament
center modulation grows rapidly with wavenumber $k_{\rm max}$, as
predicted by linear stability analysis in Eq.~\eqref{eq:7}.  Recall
that the soliton filament width is predicted to not exhibit growth
during the linear stage of evolution.  This is consistent with
Fig.~\ref{fig:NLevolution_topological}b where an initially constant
width takes some time to develop even small amplitude oscillations.
Moreover, these oscillations exhibit the wavenumber $2k_{\rm max}$,
the second harmonic of the maximally unstable mode and is due to the
nonlinear coupling of the soliton filament parameters in
Eqs.~(\ref{eq:ELave}).  As in the case of the non-topological bion
filament, the topological bion filament also exhibits break up into
two-dimensional coherent structures, signaling the breakdown of the
average Lagrangian theory.  We now investigate this regime.


\section{Break up of a bion filament}
\label{sec:bion-stripe-inst}

In this section, we perform time-dependent numerical simulations for a
bion stripe subject to small transverse perturbations. We discretize
Eqs.~\eqref{eq:Larmor} and \eqref{eq:heff} with no applied field.
Utilizing a periodic boundary, pseudo-spectral method in
space~\cite{Hoefer2012}~, we integrate in time with a fourth order
Runge-Kutta method. The domain is discretized into a mesh of
$128\times256$ gridpoints with $0.5\times0.5$ cells. The initial
condition is a static, $V=0$, bion stripe with a fundamental frequency
$|\omega_0| = (5\pi/64)^2\approx0.06$, so that the maximally unstable
wavelength, $k_\mathrm{max}$, is allowed by the grid and domain size
as the fifth Fourier mode. The topology of the bion stripe is enforced
by the profile and in-plane magnetization phase resulting from
Eqs.~\eqref{eq:biontheta} and \eqref{eq:bionphi} and the
transformation Eq.~\eqref{eq:spherical}.

\begin{figure*}
\includegraphics[width=1\linewidth]{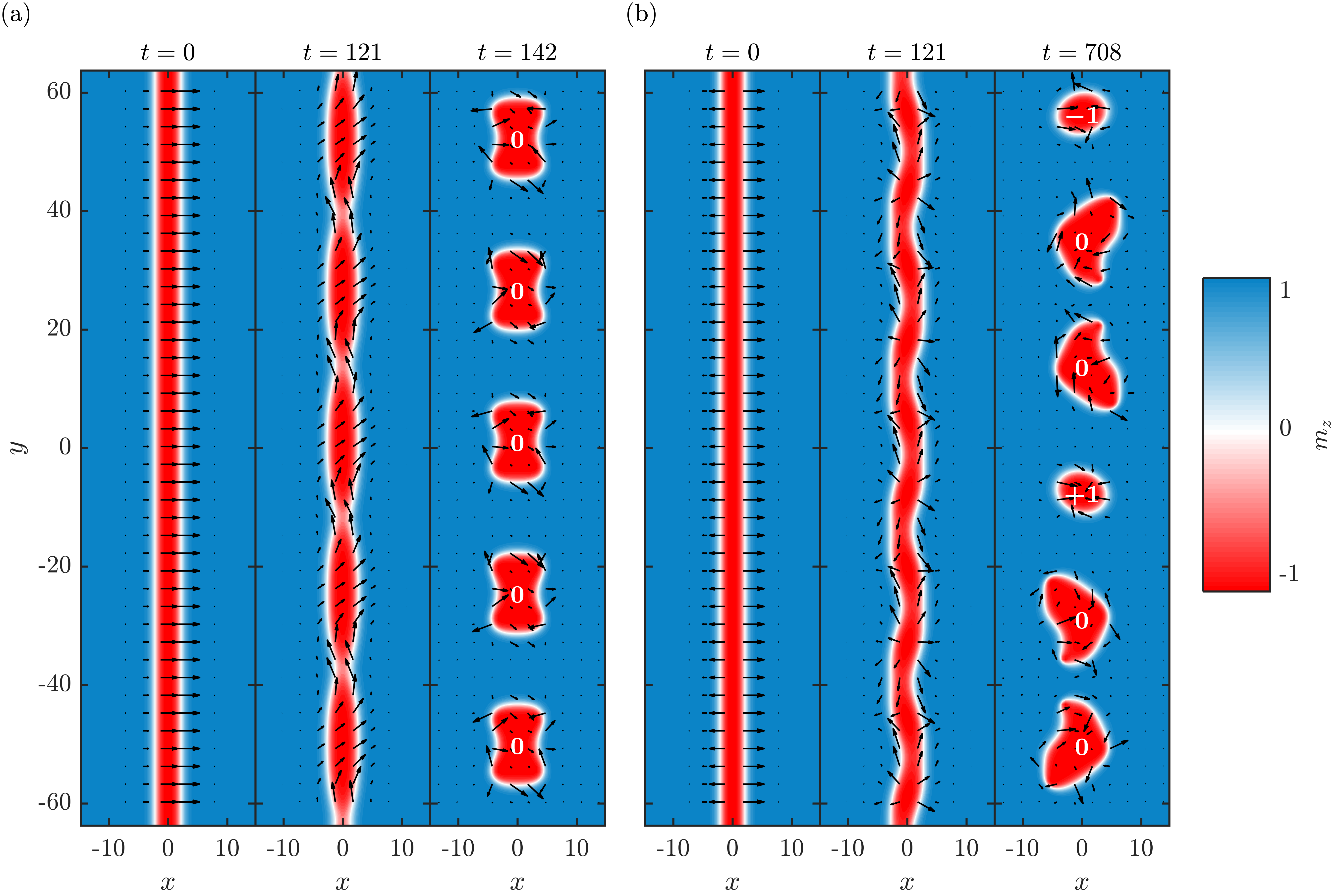}
\caption{\label{fig:evolution} (a) Evolution of the non-topological
  bion stripe, showing the neck instability. (b) Evolution of the
  topological bion stripe, showing the snake instability. The skyrmion
  number of the resultant textures are labeled in white.}
\end{figure*}

First, we consider the case of a non-topological bion stripe. The
initial condition consists of the bion stripe \eqref{eq:bion} with its
frequency modulated $\omega = \omega_0[1 + 10^{-3}\sin(k_{\rm max}
y)]$ with $k_{\rm max} = \sqrt{\omega_0} = 0.245$.  This represents a
small perturbation of the non-topological bion stripe. The temporal
evolution of the bion stripe is shown in
Fig.~\ref{fig:evolution}(a). At $t=0$, shown in the leftmost frame,
the bion stripe is only slightly perturbed in the $\hat{\mathbf{y}}$
direction. As time evolves, the unstable mode grows and develops a
symmetric, neck instability of the bion stripe width, visible in the
central panel at $t=121$ and consistent with the linear stability
analysis of Sec.~\ref{sec:bion-filam-stab}.  The observed oscillation
wavelength is 25.6, equal to the maximally unstable wavelength
$2\pi/k_{\rm max}$.  Further evolution in time deviates from the
linear regime and the bion filament pinches and is severed, leading to
the formation of two-dimensional structures. We observe five
structures that subsequently divide and merge in pairs, establishing
breathers~\cite{Maiden2013}. 
A snapshot is is shown in the rightmost frame at $t=142$. By computing
the 
skyrmion number Eq.~\eqref{eq:15} in an area that encloses each
structure and satisfies $m_z\approx1$ at the boundaries, we find
that 
the resulting structures are topologically trivial, indicating that
these breathers are composed of two droplets.

Now we consider the topological bion stripe. In this case, the
instability is favored by spatially modulating the bion's offset $\chi
= 10^{-3} \sin(k_{\rm max} y)$ where $\omega_0 = -0.06$, $k_{\rm max}
= 0.245$. The temporal evolution of the topological bion filament is
shown in Fig.~\ref{fig:evolution}(b). The leftmost frame shows the
slightly perturbed initial state at $t=0$. At time $t=121$, shown in
the central frame of Fig.~\ref{fig:evolution}(b), a snake instability
is observed. Here, the instability is dominated by a modulation of the
offset with wavelength $25.6 = 2\pi/k_{\rm max}$ but also exhibits
a slight modulation in the width. As noted in the previous sections,
this is caused by nonlinear coupling between the filament width and
center that is disregarded in the linear stability analysis considered
in Sec.~\ref{sec:bion-filam-stab}. Interestingly, the instability
leads to the separation of topological poles and anti-poles marked by
a positive or negative chirality, respectively. The time of the
central frame is the same as that of the non-topological case,
indicating that the growth rate is approximately independent of
topology in agreement with Eq.~\eqref{eq:kmax}. The poles and
anti-poles eventually separate from the bion filament and shrink below
the numerical grid scale, ultimately annihilating into a burst of spin
wave radiation.  The remains of the soliton filament establish
two-dimensional textures, as shown in the rightmost frame of
Fig.~\ref{fig:evolution}(b) at $t=708$. The dynamic evolution from the
bion filament separation to the (approximate) stabilization of
two-dimensional textures includes annihilation of topological poles
and merging of textures, requiring much longer times to stabilize. The
resultant textures in this case include a skyrmion, an anti-skyrmion,
and four droplets, conserving the system's trivial skyrmion number.

The long time
dynamics 
cannot be accurately predicted because of the interactions between
the 
resulting two-dimensional structures after the bion filamentary
breakup. However, the linear stability analysis can be used to predict
the dynamics shortly after breakup and, consequently, the number of
non-topological textures and topological poles that result from the
instability. For the non-topological bion stripe, the filament is
severed where the frequency approaches unity, i.e., a homogeneous,
out-of-plane magnetization. This implies that the number of
non-topological textures per unit length, $N_{\rm d}$, can be
estimated as the inverse of the maximally unstable wavelength
\begin{equation}
\label{eq:dropDens}
    N_{\rm d} = \frac{k_{\rm max}}{2\pi} =
    \frac{\sqrt{\omega_0}}{2\pi} .
\end{equation}
For the parameters used in the numerical simulation, $N_{\rm d} =
0.039$, so the total number of textures in a domain of height $L_y =
128$ is predicted to be $N_{\rm d} L_y = 5$. Coincidentally, this
agrees with the five breathers observed in Fig.~\ref{fig:evolution}(a)
at long times after breakup, but we stress that each breather is
composed of two droplets. For the topological bion, we can predict how
topological poles form within the bion filament. The limit $V \to
0^{\pm}$ in the phase $\Phi$ yields a $\pi$ phase jump for $\omega <
0$, as noted in Eq.~\eqref{eq:16}.  For small but nonzero $V$, we have
the expansion of Eq.~\eqref{eq:bionphi}
\begin{equation}
  \label{eq:12}
  \begin{split}
    \Phi(x,y,t) &\sim \phi -\frac{V(x-\chi)}{2} \\
    & \quad - \tan^{-1} \left [  \frac{
        F(\omega) \tanh(\sqrt{1-\omega} [x-\chi])}{V} \right ],
  \end{split}
\end{equation}
where $F(\omega) = 2 |\omega|/\sqrt{1-\omega} > 0$.  Equation
\eqref{eq:12} expresses the smoothing of the phase jump by a nonzero
velocity $V$. The phase jump is negative when $0< V \ll 1$ and
positive for $0<-V\ll 1$.  Therefore, when the velocity $V(Y,T)$
passes through zero with $V_Y < 0$ or $V_Y > 0$, a topological pole is
formed.  The sign of the pole's skyrmion number $S$ [recall
Eq.~\eqref{eq:15}] is opposite the spatial slope of $V$, i.e.,
$\mathrm{sgn} S = -\mathrm{sgn} V_y$.  Because there are two velocity
zero crossings per period of the instability, we can write the poles
per unit length as
\begin{equation}
\label{eq:poleDens}
    N_\mathrm{p} = \frac{\sqrt{\abs{\omega_0}}}{\pi}.
\end{equation}
Note that this is simply twice Eq.~\eqref{eq:dropDens}.

\section{Discussion: bion stripes in physically confined systems}
\label{sec:disc-bion-strip}

The linear stability analysis presented in
Sec.~\ref{sec:bion-filam-stab} predicts a long wavelength instability
of bion stripes and provides quantitative information on the dynamics
and nucleation of two-dimensional structures for extended thin films
under the assumption of only a local dipole field. In the case of
physically confined systems, i.e. nanowires, our approach also
provides quantitative predictions for the stabilization of bion
stripes based on the allowed wavelengths.

For a bion stripe confined in the $\hat{\mathbf{x}}$ direction, the
analysis presented above holds insofar as the bion stripe is
sufficiently localized within the nanowire. In other words,
interactions between the bion stripe and the physical boundary must be
avoided. If this condition is satisfied, bion stripes for an extended
domain in the $\hat{\mathbf{y}}$ direction will be unstable to long
wavelength perturbations and favor the growth of the maximally
unstable mode as shown in Fig.~\ref{fig:evolution}. For a bion stripe
confined along the $\hat{\mathbf{y}}$ direction with a free spin
boundary condition ($\partial_y \mathbf{m}|_{y=\pm w/2} = 0$), the
instability depends on the dimensionless width of the nanowire,
$w$. Linear stability analysis (Sec.~\ref{sec:bion-filam-stab})
predicts that a bion stripe will be stable for widths
\begin{equation}
  \label{eq:10}
  w< \frac{\pi}{k_{\rm c}} = \frac{\pi}{\sqrt{2\abs{\omega_0}}},
\end{equation}
independent of its topology. Bion stripes in wider wires are unstable
and will separate into two-dimensional textures.
The nanowire width dependency is inversely proportional
to the (square root of the) bion stripe precessional frequency. This implies that static
bion stripes or bound domain walls of either Bloch or N\'{e}el type
($\omega_0=0$) are always stable. However, perturbations from e.g.,
thermal fluctuations, can induce dynamics and modulations of the bion
filament's width and frequency. Assuming an initial bion stripe
frequency of $\omega_0=0.06$, bion stripes will be stable for nanowires
narrower than $w\approx 9$.

Physical insight on these conditions can be gained by scaling both the
bion stripe frequency $\omega_0 = 0.06$ and nanowire width $w$ to
physical units by multiplying by $\gamma\mu_0M_s(Q-1)$ and
$\lambda_\mathrm{ex}/\sqrt{Q-1}$, respectively (recall the
non-dimensionalization in Sec.~\ref{sec:analytical-model}). For
example, if we consider ultra-thin CoFeB used in
Ref.~\onlinecite{Jiang2016} with $M_s=650$~kA/m and
$K_u\approx283$~kJ/m$^3$ while assuming
$\lambda_{\mathrm{ex}}\approx6$~nm (this parameter was not
characterized in Ref.~\onlinecite{Jiang2016}), we obtain a bion stripe
precessional frequency of $90$~MHz and a maximum width for a stable
bion stripe $211$~nm. If we consider Co/Ni multilayers utilized in
Ref.~\onlinecite{Mohseni2013} with $M_s=720$~kA/m,
$K_u\approx450$~kJ/m$^3$, and $\lambda_{\mathrm{ex}}\approx8$~nm, we
obtain a bion stripe precessional frequency of $580$~MHz and a maximum
width for a stable bion stripe $117$~nm. These results are well within
state-of-the-art patterning capabilities and suggest that it is
possible to control the stability of non-topological and topological
bion stripes and their breakup into an a-priori specified/designed
number of droplets or topological poles, similar to the prescription
of the number of vortices via transverse instability in atomic
Bose-Einstein condensates~\cite{ma_2010}.  In that case too, the
complete stabilization of two-dimensional stripes in the form of dark
solitons has been advocated by suppressing the infrared catastrophes
associated with the transverse instability via increased confinement;
see, e.g., Ref.~\onlinecite{tromb_2004} and references therein.

We emphasize that these results were obtained for a minimal model of a
magnetic material with perpendicular magnetic anisotropy and local
dipole field. This assumption is justified for ultra-thin films that
can be fabricated with current deposition
methods~\cite{Jiang2016}. For relatively thick films, non-local dipole
is expected to stabilize labyrinthine domains~\cite{Hubert2009} and
further studies are needed to investigate the dynamics of bion
stripes.


\section{Conclusions}
\label{sec:conclusions}

The transverse instability of a bion stripe was studied utilizing the
average Lagrangian method.  This approach approximates the dynamics by
a soliton filament with spatio-temporally varying parameters and
complements other recent theoretical approaches to the study of
soliton filaments and
shells\cite{mironov_dynamics_2011,kevrekidis_adiabatic_2017}.  The
benefit of these approaches is a reduction in the dimensionality of
the system.  For soliton filaments in two-dimensions, their parameters
are governed by a nonlinear system of partial differential equations
in one space dimension (the transverse direction) and time.  These
equations enable an analytically tractable
linear stability analysis but their
nonlinear evolution also accurately describes the full soliton
filament's dynamics.

We found that the early stage of the transverse instability for bion
stripes can be described by exponential temporal growth in either the
filament's phase/frequency (non-topological case) or center/velocity
(topological case). We used this result to predict the most unstable
wavelength, maximal growth rate, and the dynamical manifestation of the
instability. For non-topological bion stripes, we observed a neck
instability that leads to a pinching of the bion filament.  For
topological bion stripes, we observed a snake instability that leads
to the appearance of topological poles.  Our linear stability analysis
also identifies the smallest unstable wavelength that predicts
the resulting number of two-dimensional waveforms, as well as the
complete
stabilization of bion stripes in the case of
sufficient transverse confinement.

Nonlinear evolution of the average Lagrangian modulation equations
accurately predicts the unstable dynamics up to the severing of the
bion filament.  We find that the result of filamentary breakup also
depends on topology.  For a perturbed, non-topological bion stripe, the
neck instability results in a series of two-dimensional localized
droplet solitons whose number 
can be estimated at shortly after breakup by the most unstable wavelength from linear theory. 
The snake
instability for the topological bion stripe results in a chain of
topological poles that annihilate and leave behind solitonic skyrmions, anti-skyrmions, and droplets. The topology of the
bion stripe is effectively transferred to a subset of the resultant
two-dimensional localized textures.  While the maximally unstable
wavelength from linear theory predicts the number of topological
poles, these poles are unstable and shrink to singularities that go
beyond the continuum model used here.  Further study of the dynamics
of these topological poles in a semi-classical discrete spin lattice
model may be warranted.

We note that both the bion stripes and two-dimensional textures
obtained in this study do not require chiral fields from, e.g.,
Dzyaloshinskii-Moriya interaction
(DMI)~\cite{Dzyaloshinskii1958,Moriya1960}. Bion stripes are exact
solutions of a conservative magnetic system with axial
symmetry. However, DMI may be important to stabilize bion stripes in
the presence of damping. While a study of transverse instabilities in
the presence of damping and DMI is certainly worthwhile, we here
focused on the leading energy terms that stabilize localized textures,
namely, exchange (dispersion) and anisotropy
(nonlinearity).

The average Lagrangian method applied to magnetization dynamics as
presented here can be extended to include more physics and to shed
light on the internal dynamics of droplets, skyrmions, and domain
walls.

\begin{acknowledgments}
E.I. acknowledges support from the Swedish Research Council, Reg. No. 637-2014-6863. M.A.H. partially supported by NSF CAREER DMS-1255422. M.E.R. partially supported by the National Science Foundation, DMS-1407340.
\end{acknowledgments}

\appendix

\section{Average Lagrangian for bion stripe filaments}
\label{sec:aver-lagr-bion}

The purpose of the averaged Lagrangian approach is to reduce the
dimensionality of a difficult problem, at the expense of obtaining
only an approximate description of the
dynamics\cite{malomed2002}. There are two major steps to the
approach. First, one must assume a form of the expected solution which
only explicitly depends on one of the problem's dimensions.  In the
case of bion stripes, we integrate over the moving coordinate
$\xi = x-V t$ and assume that the parameters of the bion stripe,
$\omega$, $\phi$, $V$, and $\chi$, are all functions of both the
transverse direction, $y$, and time, $t$. This assumed solution is
then substituted into the Lagrangian, Eq.~\eqref{eq:lagr}, to obtain
the Lagrangian restricted to bion filaments
\begin{equation}
  \label{eq:preL}
  L_{\rm bion} = \int_{\mathbb{R}^2} \mathcal{L}_{\rm bion}
  [\xi,\phi(y,t),\omega(y,t),\chi(y,t),V(y,t)] \mathrm{d} \xi\mathrm{d}y,
\end{equation}
where $\mathcal{L}_{\rm bion}$ is the Lagrangian density after
substitution.

The second step of the averaged Lagrangian approach is to integrate
Eq.~\eqref{eq:preL} over $\xi$. Due to the nature of the bion stripe
solution \eqref{eq:bion}, the integration can be carried out using the
Cauchy residue theorem, taking advantage of a translation symmetry in
the imaginary component of $\xi$, which is shared by many terms in
$\mathcal{L}$. While this integration is not theoretically difficult,
the Lagrangian that is obtained from the process turns out to be
complicated unless additional assumptions are made on the bion
filament's frequency $\omega$ and velocity $V$.  The obtained average
Lagrangian is therefore asymptotically expanded assuming
$0 < |\omega| \sim |V| \ll 1$.  The result, in scaled variables, is
given in Eq.~\eqref{eq:Lave}.

\section{Linearized Larmor torque equation about the bion stripe}
\label{sec:line-larm-torq}

Starting with the Larmor torque Eq.~\eqref{eq:Larmor} and
\eqref{eq:heff}, we linearize about the bion stripe solution. We
assume that the magnetization can be written as $\m = \m_0 + \m_1$,
where $\m_0$ is obtained by substituting the bion stripe solution in
Eqs.~\eqref{eq:biontheta} and \eqref{eq:bionphi}
into Eq.~\eqref{eq:spherical}. We are interested
in linearizing about the stationary bion stripe, however the
topological bion exhibits a discontinuity at $x=0$. Therefore, for
numerical stability purposes, we will consider the parameter regime
$0<\abs{V}\ll \abs{\omega}$ for the topological bion, which smooths
the discontinuity at $x=\chi(t)$ without drastically changing the
dispersion relation. For the non-topological bion, we are free to
assume $V=0$.

Because we are interested in bion stripes with a finite velocity $V$,
we transform coordinates to a moving reference frame, $\xi = x - V
t$. We can remove the explicit time dependence by 
applying a rotation matrix
\begin{equation}
\label{eq:rot}
    R(\theta) = \left[\begin{array}{ccc}
        \cos(\theta) & -\sin(\theta) & 0 \\
        \sin(\theta) & \cos(\theta) & 0 \\
        0 & 0 & 1 
    \end{array} \right]
\end{equation}
to the magnetization vector, i.e., $\m' = R(-(\omega + h_0)t)\m $. The
linearized equation for $\m'_1$ is found to be
\begin{equation}
  \label{eq:linEq}
  \begin{split}
    \partial_t \m'_1(\xi,y,t) &= V \partial_\xi \m'_1 + \omega (m'_{1,y}
    \hat{\mathbf{x}} -
    m'_{1,x}\hat{\mathbf{y}}) \\
    &- \m'_0 \times (\nabla^2 \m'_1 + m'_{1,z}
    \hat{\mathbf{z}}) \\
    &- \m'_1 \times (\partial_{\xi,\xi}\m'_0 +
    m'_{0,z} \hat{\mathbf{z}}).
  \end{split}
\end{equation}
By construction, $\m'_0$ is only a function of $\xi$, so we may assume
a linear wave solution in $y$ and $t$
\begin{equation}
\label{eq:m1Lin}
    \m'_1(\xi,y,t) = \tilde{\m}(\xi)e^{i(k y-\mu t)}.
\end{equation}
The substitution of Eq.~\eqref{eq:m1Lin} into Eq.~\eqref{eq:linEq}
yields the eigenvalue problem
\begin{equation}
  \label{eq:linEvalEq}
  \begin{split}
    \mu \tilde{\m}'_1(\xi,y,t) &= -i(V \partial_\xi \tilde{\m}'_1 +
    \omega (\tilde{m}'_{1,y}\hat{\mathbf{x}} -  \tilde{m}'_{1,x}
    \hat{\mathbf{y}}) \\
    &- \m'_0 \times (\partial_{\xi,\xi}\tilde{\m}'_1 - k^2
    \tilde{\m}'_1+ \tilde{m}'_{1,z} \hat{\mathbf{z}})\\
    &- \tilde{\m}'_1 \times (\partial_{\xi,\xi}\m'_0 + m'_{0,z}
    \hat{\mathbf{z}})).
  \end{split}
\end{equation}

We solve Eq. \eqref{eq:linEvalEq} using a numerical eigenvalue
solver. We discretize $\tilde{m}'_1$ spatially over a domain
$-11\leq \xi \leq 11$ using $10^4$ data points. This resolution is
sufficient to resolve any near-singular behavior near the origin of
the topological bion. We use second order central finite difference
stencils to estimate the derivatives in $\xi$. We impose Neumann
boundary conditions on $\tilde{\m}'_1$. The instability growth rate is
$\mathrm{Im} \mu(k)$ and the maximally unstable wavelength is found by
maximizing this function over $k$ using a numerical optimization
method.


%

\end{document}